\documentclass[10pt,a4paper,draft]{llncs}
\usepackage{amsmath,amssymb,url,color}
\DeclareMathOperator{\KS}{\textit{C}\,}
\DeclareMathOperator{\KP}{\textit{K}\,}
\DeclareMathOperator{\mm}{\mathbf{m}}
\DeclareMathOperator{\BP}{\textit{BP}\,}
\DeclareMathOperator{\BB}{\textit{BB}\,}
\newcommand{\cnd}{\mskip 1mu | \mskip 1mu}
\def\mmid{\mskip 2mu | \mskip 2mu}

\renewcommand{\le}{\leqslant}
\renewcommand{\ge}{\geqslant}
\let\eps=\varepsilon

\title{Busy beavers and Kolmogorov complexity}
\author{Mikhail Andreev}
\institute{Moscow Lomonosov State University}

\begin{document}
\maketitle

\begin{abstract}
The idea to find the ``maximal number that can be named'' can be traced back to Archimedes (see his \emph{Psammit}~\cite{arch}). From the viewpoint of computation theory the natural question is ``which number can be described by at most~$n$ bits''? This question led to the definition of the so-called ``busy beaver'' numbers (introduced by T.~Rado). In this note we consider different versions of the busy beaver-like notions defined in terms of Kolmogorov complexity. We show that these versions differ depending on the version of complexity used (plain, prefix, or a priori complexities) and find out how these notions are related, providing matching lower and upper bounds.

\end{abstract}

\section{Introduction}

In 1962 Tibor Rad\'o~\cite{rado} suggested to consider, for each natural $n$, the maximal integer that can be printed by a terminating computation of a Turing machine that has at most $n$ states. The alphabet of the machine is assumed to be binary  (blank and non-blank symbols). The machine starts on the empty tape and stops at some time. After that we count the number of non-blank symbols on the tape. Rad\'o proved that this function grows faster that any computable function (of $n$). The same is true for other functions defined in a similar way (e.g., the maximal number of steps in a terminating computation of a machine with $n$ states on the empty tape, or the maximal shift of its working head). Still these definitions look too machine-dependent: even small changes in the model (say, allowing two tapes or one-sided tape) could give different (but still fast-growing) functions. 

A more invariant approach becomes possible if we use the notions for algorithmic information theory (Kolmogorov complexity theory). We assume here that the reader is familiar with the basic notions of this theory (see, e.g.,~\cite{usv} or~\cite{livitanyi}, or the short introduction in~\cite{shen}). %Namely, we may 
We consider the maximal number that has complexity at most $n$, i.e., the maximal number that is an output of some program of length at most $n$. Here we assume that the programming language is an optimal decompressor in the sense of algorithmic information theory (that leads to a minimal complexity function; see~\cite{usv} or~\cite{livitanyi} for the formal definitions). It is easy to show (see, e.g.,~\cite[Section~1.2]{usv}) that we get the same function (up to $O(1)$-change in the argument) if we consider the maximal running time of the optimal decompressor on programs of length at most $n$. (The latter definition depends on the choice of interpreter for the optimal programming language and the computation model used to define the running time, but for every choice we get the same function up to $O(1)$-change in the argument.)

In other words, we fix optimal (plain) decompressor $D$ and denote the complexity with respect of this decompressor $D$ by $\KS(\cdot)$ (the \emph{plain Kolmogorov complexity}). Then $B(n)=\max\{N\mmid \KS(N)\le n\}$, so $B(n)$ is the maximum value of $D$ on arguments of length at most $n$ (we consider inputs as binary strings and outputs as natural numbers). Define $\BB(n)$ as the maximum computation time for $D$ on the same inputs (for arbitrary fixed machine computing $D$ in arbitrary fixed computation model). As we have mentioned, the following statement holds:
$ B(n-c)\le \BB(n)\le B(n+c) $
for some constant $c$ and for all $n$ (see~\cite{usv}).%\cite[Section~2.1]{usv}).

Additive constant in the argument is unavoidable, since the function $\KS(N)$ is defined only up to an $O(1)$ additive term (when you replace one optimal decompressor by another, an additive $O(1)$ term appears). So we will not distinguish $B(n)$ and $\BB(n)$ and will use the notation $B(n)$ in the sequel for this \emph{plain busy beaver function}.

One can repeat the same definitions for prefix-free decompressors and prefix-free Kolmogorov complexity (see~\cite{usv,shen} for the definitions). We define the \emph{prefix busy beaver function} 
$ \BP(n)=\max\{N \mmid \KP(N)\le n\}.$

Again one can consider the maximal computation time of an optimal prefix-free decompressor (as defined in~\cite[section 4.4]{usv})  on inputs of size at most $n$, and again we get two functions that are the same (up to an additive $O(1)$-term in the argument), for the same reasons.\footnote{One can define prefix complexity in different ways, using prefix-free decompressors (no element of the domain is a prefix of another element of the domain) or prefix-stable decompressors (if $D(x)$ is defined, then $D(y)=D(x)$ for every $y$ that has prefix $x$). The argument works only for prefix-free decompressors; the problem with the prefix-stable ones is that computation time of a prefix-stable decompressor is not a prefix-stable function. It would be interesting to know whether the result remains true for prefix-stable decompressors.}

So we may forget about computation time, and consider the functions $B$ and $\BP$ defined as explained above. We will compare the growth rate of the functions $B$ and $\BP$ and show that these functions are different ($B$ grows faster than $\BP$). We also compare these functions with an intermediate function $\BP'$ that will be defined in terms of the a priori probability.

Let us first recall the definition of a priori probability. A priori probability $\mm(k)$ of number $k$ can be defined as the $k$-th term of a maximal (up to multiplicative constant) converging lower semicomputable non-negative series. Levin showed that such a series exists, and proved that
$
\mm(k)=2^{-\KP(k)+O(1)}
$
(see, e.g.,~\cite[chapter 4]{usv} for the details). Now we consider the modulus of convergence for this series: for every $n$ we find minimal $N$ such that $\sum_{k>N} \mm(k) < 2^{-n}$. Denote this $N$ by $\BP'(n)$. The difference between $\BP$ and $\BP'$ can be explained as follows: after $\BP(n)$ all terms of the series $\sum \mm(k)$ are small enough (less than $2^{-n}$), and after $\BP'(n)$ the \emph{tail} of this series is small enough. Obviously $\BP(n) \le \BP'(n)$, or, more accurately, $\BP(n) \le \BP'(n + c)$ due to $O(1)$ additive terms in both definitions (Kolmogorov complexity is defined up to  an $O(1)$ additive term, and a priori probability is defined up to an $\Theta(1)$ factor).

This three functions share basic computational properties with classical %Busy Beaver
busy beaver function: they are not computable%, even more they 
and grow faster than any computable function. %Also 
All three functions are computable with 
%$0'$ 
oracle for the halting problem (as well as the classical busy beaver function.

In this article we compare growth rates %grow speed 
of these functions. Theorem~\ref{th:upper} shows that all three functions are relatively close to each other: %It implies that
all three functions are equal up to at most $(1+\eps)\log n$ argument shift. Theorem~\ref{th:gap} shows that the bound %from t
provided by Theorem~\ref{th:upper} is quite tight. % (e.g.,
For example, one cannot remove $\eps$ from the previous statement: a gap greater than $\log n$ appears between $\BP$ and $\BP'$ for some values of $n$, as well as between $\BP'$ and $B$ for some (other) values of~$n$.%).  

\begin{theorem}\label{th:upper}
\leavevmode
\textup{(i)} There exists a constant $c$ such that $\BP(n) \le \BP'(n+c)$ and $\BP'(n) \le B(n+c)$ for all $n$.

\textup{(ii)} There exists a constant $c$ such that $B(n)\le \BP(n+\KP(n)+c)$ for all $n$.

\textup{(iii)} Let $(x_n, y_n)$~be a sequence of pairs of natural numbers such that $x_n \le y_n$, the sequence $x_n$ is lower semicomputable, and the sequence $y_n$ is upper semicomputable. Assume that $\sum_n 2^{x_n - y_n} < +\infty$. Then there exists~$c$ such that $B(x_n) \le \BP(y_n + c)$ for all~$n$.

\end{theorem}

This theorem uses the notion of lower and upper semicomputable sequences. Recall that a sequence $y_n$ of real numbers is \emph{lower semicomputable} if $y_n$ is a (point-wise) limit of some total computable non-decreasing (in $k$) rational-valued function of two arguments: $y_n = \lim_k y(n,k)$; \emph{upper semicomputability} is defined in a symmetric way using non-increasing functions. If $y_n$ are natural numbers, the function $y(\cdot,\cdot)$ can be chosen in such a way that its values are also natural numbers, and convergence means that for each $n$ the equality $y_n=y(n,k)$ is true for all sufficiently large $k$. See~\cite{usv} for the details.\footnote{One may also speak about semicomputability for sequences that have terms $+\infty$ and/or $-\infty$; in this case we allow the values of the function $y(\cdot,\cdot)$ to be infinite.}

Items~(i) and~(ii) are rather simple, and~(iii) is a more symmetric way to present (ii) (as we will see later). Note that~(ii) is a special case of (iii) if we let $(x_n, y_n) = (n, n + \KP(n))$. Another special case of~(iii) is obtained if we let $(x_n, y_n) = (n - \KP(n), n)$, so $B(n - \KP(n)) \le \BP(n+c)$ for some~$c$ and all~$n$.

The statement about $(1+\eps)\log n$ mentioned above can be obtained as a corollary of~(ii) since $\KP(n)\le (1+\eps)\log n$ for $\eps > 0$ (note that the series $\sum2^{-(1+\eps)\log n}=\sum(1/n^{1+\eps})$ converges).%is converging).

Items~(ii) and~(iii) are not completely symmetric: why do we add $c$ to the right side, instead of subtracting it from the left side? We can formulate symmetric statements:

(ii$'$) $B(n - \KP(n)-c)\le \BP(n)$ for some $c$ and all $n$;

(iii$'$) Under the same assumptions as in~(iii) we have $B(x_n-c) \le \BP(y_n)$ for some~$c$ and all~$n$.

These statements are also true; we will return to them after we prove Theorem~\ref{th:upper} (they are easy corollaries of it).

\medskip

The next results say that if $\sum_n 2^{x_n-y_n} = +\infty$ (for lower semicomputable $x_n$ and upper semicomputable $y_n$) then~(iii) is not true anymore. Moreover, in this case a large gap may appear both between $B$ and $\BP'$ and between $\BP'$ and $\BP$ (but in different places).

\begin{theorem}\label{th:gap} 
Assume that $(x_n, y_n)$~is a sequence of different pairs of natural numbers, $x_n \le y_n$, the sequence $x_n$ is enumerable from below, and the sequence $y_n$ is enumerable from above. Assume also that $\sum 2^{x_n-y_n}=+\infty$. In this case

\textup{(i)} there exists $n$ such that $B(x_n) > \BP'(y_n)$;

\textup{(ii)} there exists $n$ such that $\BP'(x_n) > \BP(y_n)$.

\end{theorem}

There is no constant $c$ in this theorem (in contrast to the previous one), but one can easily put it on any side (or even both): changing all $x_n$ or all $y_n$ by an additive constant does not change the divergence condition. For example, it is true that \emph{for all $c$ there exists $n$ such that $B(x_n) > \BP'(y_n + c)$} or \emph{for all $c$ there exists $n$ such that $\BP'(x_n - c) > \BP(y_n)$}, and so on.

Using Theorems~\ref{th:upper} and~\ref{th:gap} one can easily deduce that for every upper semicomputable sequence $a_n$ the following six conditions are equivalent:
\begin{description}
\item[\textbullet] $\BP(n) \le \BP'(n+a_n+c)$ for some $c$ and for all $n$;

\item[\textbullet] $\BP'(n) \le B (n+a_n+c)$  for some $c$ and for all $n$;

\item[\textbullet] $\BP(n) \le B (n+a_n+c)$  for some $c$ and for all $n$;

\item[\textbullet] $\BP(n-a_n) \le \BP'(n+c)$  for some $c$ and for all $n$;

\item[\textbullet] $\BP'(n-a_n) \le B (n+c)$  for some $c$ and for all $n$;

\item[\textbullet] $\BP(n-a_n) \le B (n+c)$  for some $c$ and for all $n$;

\end{description}
Moreover, all these conditions are equivalent to the condition $\sum 2^{-a_n} < +\infty$ (which, in its turn, %term, 
is equivalent to $a_n\ge \KP(n)-O(1)$, see~\cite{usv}).

The meaning of Theorems~\ref{th:upper} and~\ref{th:gap} can be explained as follows. In these results we compare slow-growing functions that are \emph{inverse} to the functions $B$, $\BP$, and $\BP'$. We show that they are equal up to $(1+\eps)$ times the logarithm of their values, and that this $\eps$ cannot be omitted: without it \emph{both} inequalities between neighbor functions may be violated. As Theorem~\ref{th:upper} shows, these big gaps cannot happen at the same places (otherwise the total gap between lowest and highest functions exceeds the upper bound).

Statement~(ii) from Theorem~\ref{th:gap} has been proven by P.~G\'acs~\cite{gacs} for the case $(x_n, y_n) = (n - a_n, n)$ (and the general case may be derived as a consequence of this special one, as we will see later), so our main result is item~(i) from Theorem~\ref{th:gap}. Still we provide all the proofs in the next section for uniformity and reader's convenience.

How can we modify our definitions? One can look at the maximal $N$ such that $\KS(N\cnd n) \le n$ or such that $\KP(N\cnd n) \le n$. But we do not get new notions in this way: this quantity is still equal to $B(n)$ up to a $O(1)$-change in the argument. Indeed, the conditional complexity $\KS(x\cnd n)$ is bounded by the unconditional complexity $\KS(x)$; on the other hand, if $\KS(N\cnd n) = n$, then the conditional program of length $n$ for $N$ may be considered as a conditional prefix-free program with the same condition $n$ (if $n$ is given as a condition, we know when to stop reading the program of length $n$). Moreover, this program also can be used as unconditional program for $N$, since $n$ (its length) is determined by the program. In general, $\KP(x\cnd\KS(x))=\KS(x\cnd\KS(x))=\KS(x)$ (up to O(1) additive term), see~\cite{usv}%\cite[section 4.7]{usv}.

To finish our introduction, let us mention that $\BP'$ can be equivalently defined as the modulus of convergence for computable non-negative series of rational numbers with Martin-L\"of random sums.

\begin{theorem}
Let~$\sum a_n$ be a computable series of rational non-negative numbers whose sum is Martin-L\"of random. Let $N(\eps)$ be the modulus of convergence of this series, i.e., the minimal value of $N$ such that $\sum_{n > N} a_n < \eps$. Then $\BP'(n-c) \le N(2^{-n}) \le \BP'(n+c)$
 for some $c$ and all $n$.
\end{theorem}

The first inequality was proven in~\cite[Theorem 19]{revisited}, while the second one follows from  the definition of the a priori probability (recall that $\mm$ is bigger than any computable converging sequence, up to $O(1)$ factor). In~\cite{revisited} it was also shown that if $N(\eps)$ is the modulus of convergence for some computable converging series $\sum a_n$ with non-negative terms, and $\BP(n-c) \le N(2^{-n})$ for some $c$ and all $n$, then the same property holds for $\BP'$ (for a different value of~$c$).

\section{Upper bounds}

In this section we prove Theorem~\ref{th:upper}.

(i)~The inequality $\BP(n)\le \BP'(n+c)$ follows directly from definitions. If we define $\mm(n)$ exactly as $2^{-\KP(n)}$, it is true even without $c$-term.

Now we prove that $\BP'(n)\le B(n+c)$ for some~$c$ and for all~$n$. To do this, we construct an algorithm that, given $n$, enumerates at most $2^n$ different integers, and the last of them is bigger than $\BP'(n)$. The $n$-bit string that is the bit representation of the item's number in this enumeration, identifies the last number ($n$ is known, being the length of this string), so we get the required inequality. How the enumeration algorithm works? This algorithm approximates all $\mm(n)$ from below in parallel; we assume that at every moment only finitely many approximations are not zeros. As soon as the tail of the current approximation for~$\mm$, starting from the last enumerated integer, becomes greater than $2^{-n}$ (i.e., the current approximation to $\BP'$ exceeds the last enumerated integer) we enumerate a new integer that is bigger than all $k$ with non-zero current approximations for $\mm(k)$. Obviously this  cannot happen more than $2^n$ times: every time an integer is enumerated, we leave behind total $\mm$-weight at least $2^{-n}$.

(ii)~It is well known that $\KP(x)\le\KS(x)+\KP(\KS(x))+O(1)$ (for example, see~\cite[Section 4.6]{usv}). The following slightly more general statement is also true: if $\KS(x)\le n$, then $\KP(x)\le n+\KP(n)+O(1)$. Let us prove it. Starting with a program for $x$ that has length at most $n$, we prepend a block of the form $0^k1$ to it (this block is obviously self-delimited) making the total length exactly $n+2$. Then we prepend a self-delimited code for $n$ (of length $\KP(n)$), and the result is a self-delimited code for $x$ (decode $n$ first, then read exactly $n+2$ symbols, remove $0^k1$ leading block, then use $\KS$-decompressor). This generalisation immediately implies that $B(n) \le \BP(n+\KP(n)+c)$ for some~$c$ and for all~$n$.

(iii) We will show that this inequality is a consequence of (ii). We start by showing that we can assume $x_n$ and $y_n$ to be computable without loss of generality.

By assumption, the sequences $x_n$ [resp.~$y_n$] are lower [resp.~upper] semicomputable.  For each $n$, consider a uniformly computable sequence of pairs $(x_n^i, y_n^i)$ of integers that monotonically converge to $(x_n, y_n)$ as $i\to\infty$. Combine arbitrarily all these sequences into one sequence, leaving only the first appearance of each pair (removing all duplicates). We get a computable sequence $(\tilde{x}_i,\tilde{y}_i)$; every pair $(x_n, y_n)$ appears in this sequence together with finitely many its approximations.  Note that $\sum_i 2^{\tilde{x_i} - \tilde{y}_i}$ is at most two times bigger than $\sum_n 2^{x_n - y_n}$: every time a new approximation for $x_n$ or $y_n$ appears, the respective term is the sum is increased by factor $2$ or more, so the sum for $\tilde{x}_i, \tilde{y}_i$ is at most twice bigger than the original one, and if the original sum is finite, then the new one is also finite. Note also that the desired inequality for the new sequence implies the same inequality for the original sequence (that is a subsequence of the new one). So we can assume $x_n, y_n$ is computable without loss of generality.

Now assume that a computable sequence $(x_i, y_i)$ is given. Define 
$f(n) = \min \{y_i -n \mid x_i = n\};$
 if $n$ does not appear among $x_i$, the value $f(n)$ is $+\infty$. The function $f$ is upper semicomputable, and $\sum_n 2^{-f(n)} < +\infty$, since the pairs $(n, n+f(n))$ are guaranteed to appear among $(x_i,y_i)$.  So $f(n) \ge \KP(n) - O(1)$.  Therefore, 
 $x_i + \KP(x_i) \le y_i + O(1)$
 for the pairs with minimal $y_i$ (for a given $x_i$) and therefore for all pairs. The function~$\BP$ increases, so we get $B(x_i) \le \BP(x_i + \KP(x_i) + O(1)) \le \BP(y_i + O(1))$ for all pairs. The claim~(iii) is proven.

Symmetric results (mentioned above) are also easy to prove:

(ii$'$) $B(n-\KP(n)-c)\le \BP(n)$ for some $c$ and for all $n$.

(iii$'$) If $x_n$ and $y_n$ satisfy the same assumptions as in~(iii), then~$B(x_n-c)\le \BP(y_n)$ for some $c$ and for all~$n$.

To prove (ii$'$) we use (ii) for a smaller argument: 
$
B(n-\KP(n)-e)\le \\ \le \BP(n-\KP(n)-e+\KP(n-\KP(n)-e)+c)
$ 
holds for some $c$ and all $n, e$. Now we want to choose the constant $e$ in such a way that the argument in the right hand side is at most $n$ for all $n$ (recall that function $\BP$ is monotone): 
$n-\KP(n)-e+\KP(n-\KP(n)-e)+c \le n.$\\
Indeed, $\KP(n-\KP(n)-e)\le \KP(n,\KP(n))+\KP(e)+O(1)\le \KP(n)+\KP(e)+O(1),$
and $e-\KP(e)$ can be made arbitrary large for large enough $e$ (larger than sum of $O(1)$ terms in the inequalities).

To derive (iii$'$) from (ii$'$), one can use the same technique as used to deduce (iii) from (ii). The only difference is that one should group pairs with the same $y_i$ (instead of $x_i$, as we did in the proof).

\section{Lower bounds}

In this section we prove Theorem~\ref{th:gap}.

\subsection{Proof of the claim~(i)}

We have a sequence of different pairs $(x_n, y_n)$ of integers such that $x_n \le y_n$. We assume that $x_n$ is lower semicomputable, $y_n$ is upper semicomputable and $2^{x_n - y_n} = +\infty$. We need to show that there exists $n$ such that $B(x_n) > \BP'(y_n)$.

First we will reduce this statement to its special case where $(x_n, y_n) = (n, n + a_n)$, and $a_n$ is some upper semicomputable sequence of natural numbers (the value $+\infty$ is also allowed). 

For this reduction we use the same trick as in the previous section. First we replace $(x_n, y_n)$ by its approximations $(x_n^i, y_n^i)$, and then combine all these approximations into one computable sequence by removing the duplicates. The sum of $2^{x_i-y_i}$ may only increase (we add new elements), there are no duplicates (we removed them) and if $B(x_n^i) > \BP'(y_n^i)$ then $B(x_n) > \BP'(y_n)$ since we use monotone approximations and the busy beaver functions are monotone. So we may assume without loss of generality that the sequence $(x_n, y_n)$ is  a computable sequence of different integer pairs.

Let $a_n = \min \{y_i - n \mid x_i = n\}$. The sequence $a_n$ is enumerable from above (since the sequence $(x_i, y_i)$ is computable). Note that $\sum_n 2^{-a_n} \ge \frac{1}{2}\sum_i 2^{x_i - y_i}$. Indeed, if we group pairs with $x_i = n$, the sum of this group is bounded by a geometric sequence with common ratio $1/2$, so the sum can be replaced by the maximal element (up to a $2$-factor). Therefore, $\sum_n 2^{-a_n} = +\infty$, and all pairs $(n, n+a_n)$ appear among $(x_i, y_i)$, so we get the desired reduction.

Now we use the following lemma: \emph{if $a_n$ is an upper semicomputable sequence of integers and $\sum_n 2^{-a_n}=+\infty$, there exists a computable sequence $\tilde{a}_n\ge a_n$ such that $\sum_n 2^{-\tilde{a}_n}=+\infty$.} Indeed, we can approximate $a_n$ from above until some finite part of the series $\sum 2^{-a_n}$ exceeds $1$, then fix the current approximations for this part and call them $\tilde{a}_n$. Then the same argument is used for the tail, etc. This argument show that we may assume without loss of generality that $a_n$ is a computable sequence.

It remains to prove the following statement: if $a_n$ is a computable sequence of integers and $\sum 2^{-a_n}=+\infty$, then there exists $n$ such that $B(n) > \BP'(n+a_n)$. In other words, we need to show that \emph{there exists some $u$ such that $\KS(u)\le n$ and $\sum_{i\ge u} \mm(i)<2^{-n-a_n}$}.

To prove an upper bound for $\KS(u)$, we need to construct a decompressor that provides a short description for $u$. However, this gives a bound with some additive constant term, so we need to construct a decompressor $D$ such that \emph{for every $d$ there exist $n$ and $u$ such that}
$$
\KS_D(u)\le n-d \quad \text{and}\quad \sum_{i\ge u} \mm(i)< 2^{n-a_n}.
$$
where $\KS_D(u)$ is the minimal length of $p$ such that $D(p)=u$.

To prove this, we use the game technique. Consider a game where Alice plays with Bob. They make alternating moves. Alice enumerates sets $D_0,D_1,\ldots$; at each move she adds finitely many integers to finitely many $D_i$ (so her move is a finite object). The set $D_i$ may contain at most $2^i$ elements. Bob approximates from below some sequence $\mu(0),\mu(1),\ldots$; initially all $\mu(i)$ are zeros, and at each step Bob may increase finitely many of them by some rational numbers, but the sum $\sum \mu(i)$ should not exceed~$1$.

Assuming that both players respect the rules, Alice wins if (for limit values of $D_i$ and $\mu(i)$) \emph{for every $d$ there exists $n$ and $u$ such that $u\in D_{n-d}$ and $\sum_{i\ge u} \mu(i)<2^{-n-a_n}$}. One may reformulate this statement eliminating $u$: for every $d$ there exist $n$ such that 
$$
\sum_{i \ge \max(D_{n-d})}\mu(i)< 2^{-n-a_n}.\eqno(*)
$$

We will prove that Alice has a computable winning strategy in this game. This implies the desired result. Indeed, we may let Alice use this strategy against the ``blind'' strategy of Bob that approximates from below the a priori probability function $\mu(i)=\mm(i)$. Then the behavior of Alice is computable, the sets $D_i$ are enumerable and we construct a decompressor $D$ that maps $k$-bit string $p$ into $p$th element in the enumeration of $D_k$ (in the last sentence binary string $p$ is identified with an integer it represents in the binary notation). This decompressor has the required property.

So why Alice has a computable strategy? She should guarantee the existence of a suitable $n$ for each $d$. This is done independently for each $d$; Alice chooses for each $d$ some interval $[l_d,r_d]$ where $n$ with the required properties exist. This intervals are chosen in such a way that there are no collisions (for different $d$ the values of $n-d$ cannot be the same, i.e., the intervals $[l_d-d,r_d-d]$ are disjoint). The intervals should be large enough: the sum of $2^{-a_n}$ over $n$ in $[l_d,r_d]$ should exceed $2^{d+1}$ (we will see that this is enough for our purposes). Since we assume that $a_n$ is a computable sequence and $\sum 2^{-a_n}=+\infty$, we can choose $[l_d,r_d]$ in a computable way.

How Alice constructs $D_{n-d}$ for $n\in[l_d,r_d]$? It is done in a straightforward way. Alice chooses some $n$ (say, the minimal value $n=l_d$) and tries to achieve $(*)$ by adding large elements to $D_{n-d}$. More precisely, if $(*)$ is violated, Alice takes some number $k$ that is greater that all non-zero terms in $\mu$ (i.e., $\mu(k')=0$ for all $k'\ge k$) and adds $k$ to $D_{n-d}$. Then Bob may increase $\mu$-values; as soon as $(*)$ is violated again, Alice repeats this procedure, and so on. At some point (after $2^{n-d}$ steps) a maximal cardinality of $D_{n-d}$ is reached. But at that time Bob has used at least $2^{n-d}2^{-n-a_n}=2^{-d-a_n}$ of his reserve (each time a tail of size $2^{-n-a_n}$ is cut). Then Alice switches to next value of $n$, and forces Bob to lose or to use $2^{-d-a_n}$ again for this new value of $n$. Ultimately Bob will lose the game since the sum of $2^{-d-a_n}=2^{-d} 2^{-a_n}$ over $n$ in $[l_d,r_d]$ exceeds $1$. (A technical correction: we required that the limit value of $\sum \mu(i)$ is strictly less that some threshold; it is not enough to know that all the approximations are strictly less than this threshold (only a non-strict %non-string 
inequality is guaranteed). To remedy this problem, we may use an additional factor of $2$ --- so we require the sum of $2^{-a_n}$ over $n\in [l_d,r_d]$ to be greater than $2^{d+1}$, not $2^d$.) Claim~(i) is proven.

\subsection{Proof of the claim~(ii)}

We again consider a sequence of different pairs $(x_n, y_n)$ %%%%, 
such that $x_n \le y_n$, the sequence $x_n$ is lower semicomputable, the sequence $y_n$ is upper semicomputable and $\sum 2^{x_n - y_n} = +\infty$. We want to prove (following~G\'acs) that there exists $n$ such that $\BP' (x_n) > \BP (y_n)$

We can use the same reasoning as in~(i) with minor modifications to show that we can assume without loss of generality that $(x_n, y_n) = (n - a_n, n)$  for some computable sequence of non-negative integers $a_n$ 
with $\sum_n 2^{-a_n} = +\infty$. This time we need to group terms with the same $y_i$, not $x_i$. We need to prove then that there exists $n$ such that $\BP(n) < \BP'(n-a_n)$. In other words, we need to prove that there exist $n$ and $u$ such that $\mm(i)<2^{-n}$ for all $i\ge u$, but $\sum_{i\ge u} \mm(i)>2^{-n+a_n}$ (all terms in the $u$-tail are small but their sum is big).

To show that the sum of $\mm$-tail is big, we need to construct a lower semicomputable semimeasure for which this sum is big, and then use the maximality of $\mm$. Again a constant appears, so we need to prove a stronger statement: \emph{there exists a lower semicomputable semimeasure $\alpha$ such that for every $d$ there are $n$ and $u$ with the following property:}
$$
\sum_{i\ge u} \alpha(i)> 2^{-n+a_n+d} 
\quad
\text{but}
\quad
\mm (i)<2^{-n} \text{ for all } i\ge u.
$$
Again we may use the game approach and imagine that Alice approximates from below some semimeasure $\alpha$ while Bob approximates from below some semimeasure $\beta$, and the claim above (with $\beta$ instead of $\mm$) is the winning condition for Alice. We will construct a computable strategy for Alice in this game; applying it against the blind strategy of Bob (who approximates $\mm(\cdot)$ from below), we get the required statement.

Let us note first that it is enough to construct (for every $d$) a winning strategy in the similar game where winning condition is required only for this $d$. Indeed, we may use $2d$-strategy to win the $d$-game with $\sum_i \alpha(i)\le 2^{-d}$ (using $2d$-strategy with factor $2^{-d}$). Then we can use all the strategies (for $d$-games for all $d$) in parallel against Bob and sum up all the increases, since the winning condition is monotone and the strategies can only help each other. In this way Alice keeps the total sum less than $\sum_d 2^{-d}\le 1$ and wins all games.

So how could Alice win the $d$-game? She should increase her weights gradually by using small weights far away where Bob has only zeros. As soon as her total weight exceeds $2^{-n+a_n+d}$ for some $n$, Bob has to react and assign weight at least $2^{-n}$ for some $i$.  Then Alice continues to increase the weights (on the right of the place used by Bob), and again after $2^{-n+a_n+d}$ new Alice's weight Bob should react by assigning weight at least $2^{-n}$ at some other place. If Alice uses this strategy with small weights (see the discussion below) until her total weight reaches $1$, and waits each time until Bob violates the winning condition for Alice, we have the following property of Bob's weights:\footnote{Technically speaking, Bob is obliged to react only if the Alice's tail is strictly greater than $2^{-n+a_n+d}$. But this leads only to a constant factor that is not important, so we ignore this problem.}
\begin{quote}
\emph{for each $n$ there are at least $2^{n-a_n-d}$ Bob's weights $\beta(\cdot)$ that exceed $2^{-n}$.}
\end{quote}
Note that Alice's actions are the same for all $n$; it is Bob who should care about all $n$ and provide a large enough weight at the moments where Alice is in the winning position (for some~$n$).

What is the total weight Bob uses in this process? The property above guarantees that Bob uses at least $2^{-a_n-d}$ to prevent Alice from winning for given $n$. The sum of these quantities for all $n$ is infinite according to our assumption (so at some point Bob will be unable to increase the weights). However, the same Bob's move can be useful on different levels (for different values of $n$), so we need the following technical lemma valid for every series $\sum_i \beta(i)$ with non-negative values:
   $$\sum_j 2^{-j}\cdot \#\{i: \beta(i) \ge 2^{-j}\} \leqslant 2\sum_i \beta(i).$$
Indeed, each $\beta(i)$ from the right hand side appears in the left hand side as the sum of $2^{-j}$ for all $j$ such that $2^{-j} \le \beta(i)$, and this sum does not exceed $2\beta(i)$.

To finish the description of Alice's strategy, we need to say how small should be the weight increases %in weight 
used by Alice. We know that the sum $\sum_n 2^{-a_n-d}$ is infinite, so there is a finite part of this sum that is large (greater than $4$, to be exact). Alice then may use weights $2^{-s}$ where $s$ is some integer greater that all $n+a_n+d$ for $n$ that appear in this finite part.

\section{Acknowledgements}
This work was supported by  ANR-15-CE40-0016-01 RaCAF grant, and RFBR grant number 16-01-00362.

\end{document}